\documentclass[aps,prb,a4paper,10pt,twocolumn,showpacs,floatfix,superscriptaddress,preprintnumbers,longbibliography]{revtex4-2}
\usepackage{float}
\usepackage{dcolumn,graphicx,color,booktabs,microtype,afterpage}
\usepackage{amssymb}
\usepackage{amsmath}
\usepackage[charter,greekuppercase=italicized]{mathdesign}
\usepackage{sidecap}
\usepackage[mathlines]{lineno}

\graphicspath{{./}{figure/}}
\renewcommand{\tablename}{Table}
\makeatletter\renewcommand{\fnum@figure}[1]{\figurename~\thefigure.~}\makeatother
\makeatletter\renewcommand{\fnum@table}[1]{\tablename~\thetable.}\makeatother

\newcount\hh \newcount\mm
\hh=\time \divide\hh by 60
\mm=\hh \multiply\mm by 60 \mm=-\mm
\advance\mm by \time
\def\now{\number\hh:\ifnum\mm<10{}0\fi\number\mm}

\usepackage[colorlinks,plainpages=false,linkcolor=blue,urlcolor=blue,citecolor=blue,pdfpagemode=UseNone,pdfstartview=FitBH]{hyperref}

\begin{document}

\makeatletter\renewcommand{\ps@plain}{%
\def\@evenhead{\hfill\itshape\rightmark}%
\def\@oddhead{\itshape\leftmark\hfill}%
\renewcommand{\@evenfoot}{\hfill\small{--~\thepage~--}\hfill}%
\renewcommand{\@oddfoot}{\hfill\small{--~\thepage~--}\hfill}%
}\makeatother\pagestyle{plain}

\preprint{\textit{Preprint: \today, \now.}} %For internal use only, do not distribute.}}
%\linenumbers

\title{Spin order and dynamics in the topological rare-earth germanide semimetals}

\author{Yuhao Wang}\thanks{These authors contributed equally}
\affiliation{Key Laboratory of Polar Materials and Devices (MOE), School of Physics and Electronic Science, East China Normal University, Shanghai 200241, China}
\author{Zhixuan Zhen}\thanks{These authors contributed equally}
\affiliation{Key Laboratory of Polar Materials and Devices (MOE), School of Physics and Electronic Science, East China Normal University, Shanghai 200241, China}
\author{Jing Meng}
\affiliation{Key Laboratory of Polar Materials and Devices (MOE), School of Physics and Electronic Science, East China Normal University, Shanghai 200241, China}
\author{Igor Plokhikh}
\affiliation{Laboratory for Multiscale Materials and Experiments, Paul Scherrer Institut, Villigen PSI CH-5232, Switzerland}
\author{Delong Wu}
\affiliation{Key Laboratory of Polar Materials and Devices (MOE), School of Physics and Electronic Science, East China Normal University, Shanghai 200241, China} 
\author{Dariusz J. Gawryluk}
\affiliation{Laboratory for Multiscale Materials and Experiments, Paul Scherrer Institut, Villigen PSI CH-5232, Switzerland}
\author{\\Yang Xu}
\affiliation{Key Laboratory of Polar Materials and Devices (MOE), School of Physics and Electronic Science, East China Normal University, Shanghai 200241, China}
\author{Qingfeng Zhan}
\affiliation{Key Laboratory of Polar Materials and Devices (MOE), School of Physics and Electronic Science, East China Normal University, Shanghai 200241, China}
\author{Ming Shi}
\affiliation{Center for Correlated Matter and School of Physics, Zhejiang University, Hangzhou 310058, China}
\author{Ekaterina Pomjakushina}\email[Corresponding authors:\\]{ekaterina.pomjakushina@psi.ch}
\affiliation{Laboratory for Multiscale Materials and Experiments, Paul Scherrer Institut, Villigen PSI CH-5232, Switzerland}
\author{Toni Shiroka}\email[Corresponding authors:\\]{tshiroka@phys.ethz.ch}
\affiliation{Laboratory for Muon-Spin Spectroscopy, Paul Scherrer Institut, Villigen PSI CH-5232, Switzerland}
\affiliation{Laboratorium f\"ur Festk\"orperphysik, ETH-H\"onggerberg, Z\"urich CH-8093, Switzerland}
\author{Tian Shang}\email[Corresponding authors:\\]{tshang@phy.ecnu.edu.cn}
\affiliation{Key Laboratory of Polar Materials and Devices (MOE), School of Physics and Electronic Science, East China Normal University, Shanghai 200241, China}
%
%\author{J.\ Mesot}
%\affiliation{Laboratorium f\"ur Festk\"orperphysik, ETH Z\"urich, CH-8093 Zurich, Switzerland}
%\affiliation{Paul Scherrer Institut, CH-5232 Villigen PSI, Switzerland}
%

\begin{abstract}
The $RE$Al(Si,Ge) ($RE$ = rare earth) family, known to break both the inversion- and time-reversal symmetries, represents one of the most suitable platforms for investigating the interplay between correlated-electron phenomena and topologically nontrivial bands. Here, we report on systematic magnetic, transport, and muon-spin rotation and relaxation ({\textmu}SR) measurements on (Nd,Sm)AlGe single crystals, which exhibit antiferromagnetic (AFM) transitions at $T_\mathrm{N} = 6.1$ and 5.9\,K, respectively. In addition, NdAlGe undergoes also an incommensurate-to-commensurate ferrimagnetic transition at 4.5\,K. Weak transverse-field {\textmu}SR measurements confirm
the AFM transitions, featuring a $\sim$90\% magnetic volume fraction. In both cases, zero-field (ZF) {\textmu}SR measurements 
reveal a more disordered internal field distribution in NdAlGe than in SmAlGe, reflected in a larger transverse muon-spin relaxation rate $\lambda^\mathrm{T}$ at $T \ll T_\mathrm{N}$. This may be due to the complex magnetic structure 
of NdAlGe, which undergoes a series of metamagnetic transitions in an external magnetic field, while SmAlGe shows only a robust AFM order. In NdAlGe, the topological Hall effect (THE) appears between the first
and the second metamagnetic transitions for $H \parallel c$, while it is absent in SmAlGe. 
Such THE in NdAlGe is most likely attributed to the field-induced topological spin textures. The longitudinal muon-spin relaxation rate $\lambda^\mathrm{L}(T)$, diverges near the 
AFM order, followed by a clear drop at $T < T_\mathrm{N}$. In the magnetically ordered state, spin fluctuations are significantly stronger in NdAlGe than in SmAlGe.
In general, our longitudinal-field {\textmu}SR data indicate vigorous spin fluctuations in NdAlGe, thus providing valuable insights into the origin of THE and of the possible topological spin textures in $RE$Al(Si,Ge) Weyl semimetals.
\end{abstract}

%\keywords{Unconventional supeconductivity, muon spin rotation/relaxation, nuclear magnetic resonance}

\maketitle\enlargethispage{3pt}

\vspace{-0pt}
\section{\label{sec:Introduction}Introduction}\enlargethispage{8pt}
In recent years, the magnetic topological materials have attracted
extensive research interests due to the interplay between correlated-electron
phenomena and topological electronic bands~\cite{lv2021,bernevig2022,yan2012}.
Weyl semimetals represent one of the most interesting subclasses
of topological materials. They are characterized by linearly dispersed
electronic bands and often exhibit large Berry curvatures near the
Fermi surface~\cite{yan2017,wang2017,armitage2018}. 
The Weyl-semimetal phase emergences only upon breaking the space
inversion or the time-reversal symmetry, the latter being realized
either by applying an external magnetic field, or via
intrinsic magnetic order~\cite{yan2017,wang2017,armitage2018,liu2019}.
When both symmetries are broken, as is the case of noncentrosymmetric
magnetic materials, the topological electronic bands are strongly coupled
to the magnetism. As a consequence, the Weyl nodes and the associated
Berry curvatures can be tuned by various external parameters, as e.g.,
the magnetic field~\cite{puphal2020,ueda2023}, the chemical
substitution~\cite{puphal2020-1,fujishiro2019}, the physical
pressure~\cite{piva2023-1,sun2021-1}, or the epitaxial
strain~\cite{ohtsuki2019}. Such tuning often results in a series
of exotic properties, such as the topological Hall effect (THE) or
the anomalous Hall effect (AHE)~\cite{xu2021,yao2023}.   

To date, Weyl semimetals with both broken inversion and
time-reversal symmetries are rare. The $RE$Al(Si,Ge) ($RE$ = rare earth)
family of silicides or germanides represents one of the ideal platforms
to investigate the tunable interplay between Weyl physics and magnetism.
Depending on the rare-earth radius, the $RE$Al(Si,Ge) family adopts
either a noncentrosymmetric tetragonal \
structure with a space group $I4_1md$ (No.~109) (for low-$Z$,
large radius, i.e., La--Eu), or a centrosymmetric orthorhombic 
structure with a space group $Cmcm$ (No.~63) (for high-$Z$, small
radius, i.e., Gd--Lu)~\cite{gouda2022,Wang2021,Qin2011,Pukas2004}.
In the former group, only a few compounds have been predicted or
experimentally demonstrated to be Weyl semimetals~\cite{chang2018,xu2017,sanchez2020,su2021,zhang2023,li2023}.
The noncentrosymmetric family also exhibits a variety of exotic
properties, including pressure-induced superconductivity in LaAlGe~\cite{cao2022}, topological spin textures and THE in CeAlGe~\cite{puphal2020,He2023}, and a large AHE in PrAlGe~\cite{meng2019,destraz2020}. The breaking of inversion symmetry can give rise to Dzyaloshinskii-Moriya interactions (DMIs) which, through the
competition with the Ruderman-Kittel-Kasuya-Yosida (RKKY) interactions,
leads to rich spin structures in the $RE$Al(Si,Ge) family. 
Notable cases include the topological multi-$\boldsymbol{k}$ structure in CeAlGe~\cite{puphal2020}, the spiral order in SmAlSi~\cite{yao2023}, and the incommensurate spin-density-wave (SDW) order in NdAl(Si,Ge)~\cite{Gaudet2021,Dhital2023}.

Interestingly, upon further lowering the temperature, NdAlGe
undergoes another magnetic transition, from the incommensurate
SDW state to a commensurate ferrimagnetic (FIM) state~\cite{Dhital2023,Yang2023}.
In both magnetically ordered states, NdAlGe shows a large AHE, mostly attributed
to the significant Berry curvatures near the Fermi surface~\cite{Cho2023}. 
In case of a sizable disorder, also extrinsic mechanisms, 
including skew scattering or side-jump, may contribute to the AHE in
NdAlGe~\cite{Yang2023}. When applying a magnetic field along
the $c$ axis, NdAlGe undergoes a series of metamagnetic 
transitions~\cite{Dhital2023,Yang2023}. Within the 0 to 9\,T field range,
its field-dependent Hall resistivity $\rho_{xy}(H)$ closely resembles that of the sister compound
CeAlGe~\cite{puphal2020}. After subtracting the ordinary
and the anomalous contributions from the $\rho_{xy}(H)$ of CeAlGe, the topological contribution could be isolated, here
induced by a multi-$\boldsymbol{k}$ magnetic
structure~\cite{puphal2020}.
Although signatures of topological Hall resistivity have been
observed in NdAlGe~\cite{kikugawa2024}, its field and temperature dependence,
as well as its origin remain largely unexplored. Moreover, to date,
most of the studies have been targeting $RE$AlGe ($RE$ = La, Ce, Pr, and Nd) and much less is known about SmAlGe.

Here, we report on the systematic magnetization, electrical
resistivity, Hall resistivity, and muon-spin rotation and relaxation ({\textmu}SR) measurements on $RE$AlGe ($RE$ = Nd, Sm)
single crystals. Our extensive datasets allowed us to establish
their detailed magnetic phase diagrams. While NdAlGe shows clear
evidence of anomalous and topological Hall resistivity, the ordinary Hall resistivity
is dominant in SmAlGe. In the following, we discuss the possible reasons
for the occurrence of THE in NdAlGe and its absence in SmAlGe. In both cases,
our {\textmu}SR measurements reveal
significant and robust spin fluctuations, which might be crucial
to the formation of THE and of topological spin textures in $RE$Al(Si,Ge) Weyl semimetals.

\section{Experimental details\label{sec:details}}\enlargethispage{8pt}
Single crystals of NdAlGe and SmAlGe were grown using a molten Al-flux
method~\cite{Puphal2019}, with the remaining Al flux removed by centrifugation at 750$^\circ$C after the synthesis.
Crystal orientation and phase purity were checked by x-ray diffraction (XRD) measurements using a Bruker D8 diffractometer with Cu K$\alpha$ radiation. The atomic ratios of the NdAlGe and SmAlGe single crystals were measured by x-ray fluorescence spectroscopy (XRF) on an AMETEK Orbis Micro-XRF analyzer. Crystal structure was further checked by single-crystal x-ray diffraction measurements performed at 120\,K using  a STOE STADIVARI diffractometer with Mo K$\alpha$ radiation.
%Single crystal x-ray diffraction measurements were performed at 120 K using the STOE STADIVARI diffractometer equipped with a Dectris EIGER 1M 2R CdTe detector and with an Anton Paar Primux 50 Ag/Mo dual-source using Mo$\mathrm{K_\alpha}$ radiation ($\lambda$ = 0.71047 $\AA{}$) from a micro-focus X-ray source and coupled with an Oxford Instruments Cryostream 800 jet.} 
The electrical transport and magnetic properties were measured in a Quantum Design physical property measurement system and magnetic properties measurement system, respectively. For the above measurements, the magnetic field was applied both along the $c$ axis ($H \parallel c$)
and within the $ab$ plane (i.e., $H \parallel ab$). Also for the
transport measurements, the electric current was applied both along
the $c$ axis ($I \parallel c$) and within the $ab$ plane ($I \parallel ab$). For the transport measurements, a four-probe method with a dc current of 5\,mA was used.
To avoid spurious transport contributions due to misaligned Hall contacts,
the longitudinal contribution to the Hall resistivity $\rho_{xy}$, was
removed by an anti-symmetrization procedure, i.e.,
$\rho_{xy}(H) = [\rho_{xy}(H) - \rho_{xy}(-H)]/2$.
Whereas, in the case of the longitudinal electrical resistivity 
$\rho_{xx}$, the spurious transverse contribution was removed by
a symmetrization procedure, i.e., $\rho_{xx}(H) = [\rho_{xx}(H) + \rho_{xx}(-H)]/2$.

The {\textmu}SR measurements were carried out at the general-purpose surface-muon (GPS) instrument at the $\pi$M3 beam line of the
Swiss muon source (S$\mu$S) at Paul Scherrer Institut (PSI) in Villigen, Switzerland.
The aligned $RE$AlGe single crystals were positioned on a copper plate using diluted GE varnish, with their $c$ axis parallel to the muon momentum direction, i.e., $\boldsymbol{P}_\mu$ $\parallel$ $c$. 
In the current study, we performed three kinds of experiments: weak transverse-field (wTF), zero-field (ZF), and longitudinal-field (LF) {\textmu}SR measurements.
As to the former technique, we used it to determine the temperature evolution of the magnetic
volume fraction and the ordering temperature. As to the latter two,
we aimed at investigating the magnetically ordered phase and the dynamics of spin fluctuations at different temperatures. All the {\textmu}SR spectra were collected upon heating the crystals and were analyzed by means of the \texttt{musrfit} software package~\cite{Suter2012}.

%\\
%\vfill
%\vspace{\fill}
%\\
%{\par\vspace{\fill}} %{\newpage} {\clearpage}
%\columnbreak
%\newpage

\section{\label{sec:results}Results and discussion}\enlargethispage{8pt} 
\subsection{\label{ssec:structure}Crystal structure}
%%%%%%%%%%%%%%%%%%%%%%%%%
\begin{figure}[!thp]
	\centering
	\includegraphics[width=0.46\textwidth]{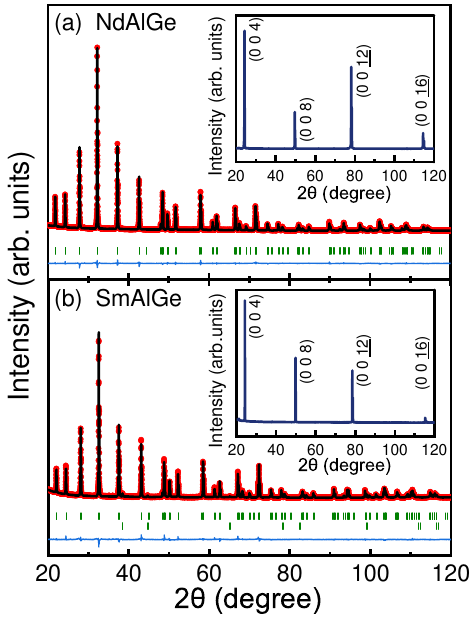}
	\caption{(Color online) Room-temperature x-ray powder diffraction
		pattern and Rietveld refinements for NdAlGe (a) and SmAlGe (b).
		The red circles and the solid black line represent the experimental pattern and the Rietveld refinement profile, respectively. The blue line at the bottom shows the residuals, i.e., the difference between the calculated and experimental data. The vertical green bars mark the calculated Bragg-peak positions. Note that, for SmAlGe, a tiny amount of Al phase ($\sim$6\%) comes from the remaining flux on the surface of crystals [see second row of Bragg peaks in panel (b)]. The insets show the XRD patterns collected on $RE$AlGe single crystals, which confirm their (00$l$) orientation.} 
	\label{fig:1}
\end{figure}
%%%%%%%%%%%%%%%%%%%%%%%%%
%
%%%%%%%%%%%%%%%%%%%%%%%%%%%%
\begin{figure}[!htp]
	\centering
	\includegraphics[width=0.46\textwidth]{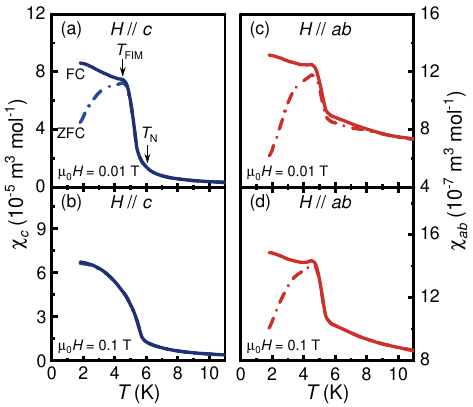}
	\caption{(Color online) Temperature-dependent ZFC and FC magnetic susceptibility $\chi(T)$ collected in a magnetic field of 0.01\,T (a) and 0.1\,T (b), applied along the $c$ axis of a NdAlGe single crystal. The analogous results with the magnetic field applied within the $ab$ plane are shown in panels (c) and (d), respectively. The solid and dash-dotted lines represent the $\chi(T)$ collected in ZFC and FC conditions, respectively. In panel (b), the ZFC and FC susceptibilities overlap.
		The arrows in panel (a) mark the magnetic transitions at $T_\mathrm{N}$ and $T_\mathrm{FIM}$.}
	\label{fig:2}
\end{figure} 
%%%%%%%%%%%%%%%%%%%%%%%%%%

The obtained $RE$AlGe single crystals were first checked by powder XRD
at room temperature.
As shown in Fig.~\ref{fig:1}, the Rietveld refinements of the powder XRD pattern were performed via the  \texttt{FullProf} suite~\cite{Carvajal1993}. Consistent with previous
studies~\cite{Dhital2023,Yang2023}, our $RE$AlGe crystals, too, exhibit a LaPtSi-type noncentrosymmetric tetragonal structure with a space group $I4_1md$ (No.~109).  
The refined lattice parameters are $a = 4.2355(1)$~\AA{} and $c = 14.6568(1)$~\AA{} for NdAlGe, and $a = 4.1933(1)$~\AA{} and $c = 14.5733(1)$~\AA{} for SmAlGe, the former being in good agreement with the results reported in the literature~\cite{Dhital2023,Yang2023}.
By replacing the Nd atoms with Sm, the in-plane and out-of-plane
lattice parameters are reduced by a factor of 1.2\% and 0.5\%, respectively
(a consequence of the lanthanide contraction). The noncentrosymmetric tetragonal crystal structure of NdAlGe and SmAlGe was further confirmed by single-crystal x-ray diffraction (see details in Supplementary Materials). 
According to the analysis, 4$a$ positions of Nd/Sm and Al atoms are fully occupied by the nominal elements, whereas the 4$a$ position of Ge atoms is interpreted as mixed Ge/Al with an Al fraction of 12.5(5)\% and 4.5(8)\% in NdAlGe and SmAlGe, respectively. 
%in NdAlGe single crystal, and 4.5(8)\% in SmAlGe case. 
This leads to compositions NdAl$_{1.13}$Ge$_{0.88}$ and SmAl$_{1.05}$Ge$_{0.96}$, consistent 
%The results suggest that several Ge sites are occupied by Al, which is consistent with the XRF chemical analysis with Nd:Al:Ge = 1:0.95:0.84 and Sm:Al:Ge = 1:1.1:0.95 for NdAlGe and SmAlGe crystals, respectively.
with the XRF analysis, giving a chemical stoichiometry of NdAl$_{0.95}$Ge$_{0.84}$
and SmAl$_{1.1}$Ge$_{0.95}$. Therefore, both NdAlGe 
and SmAlGe grown using the Al-flux method are rich in Al, unlike the Ge-rich crystals grown by the floating-zone method~\cite{Kikugawa2023}. %\tcr{\sout{Since the magnetic properties of NdAlGe are significantly different from those of SmAlGe (see below), this indicates the high sensitivity of magnetic interactions in $RE$AlGe compounds  to the local environment. Therefore, the magnetic interactions can be easily tuned by modifying external parameters such as the pressure or magnetic field.}} 
According to the refinements, while NdAlGe shows a pure phase, a tiny amount of elemental aluminum ($\sim$6\%) was identified in SmAlGe, here attributed to the remanent Al flux on the surface of single crystals.

Considering the small amount of Al and that its superconducting critical field is low, the presence of aluminum has negligible effects on the magnetic properties of $RE$AlGe. 
In any case, the resistivity and magnetization data were collected on
polished single crystals, where the Al flux on their surfaces was removed beforehand.  
We also performed XRD measurements on $RE$AlGe single crystals. As shown in the insets of Fig.~\ref{fig:1}, only the (00$l$) reflections were detected, confirming that the $c$ axis is perpendicular to the measured crystal plane.

\subsection{Magnetic and transport properties of NdAlGe}\label{sec:NAG}
%%%%%%%%%%%%%%%%%%%%%%
\begin{figure}[!htp]
	\centering
	\includegraphics[width=0.46\textwidth]{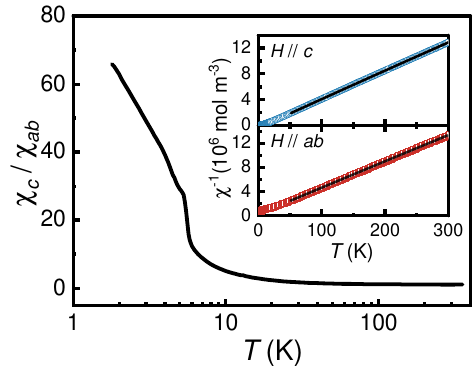}
	\caption{(Color online) Temperature dependence of the magnetic an\-iso\-tro\-py of a NdAlGe single crystal,
		here reflected by the $\chi_c$/$\chi_{ab}$ ratio. The $\chi_c(T)$ and $\chi_{ab}(T)$ datasets were collected in a field of 0.1\,T after zero-field cooling. 
		The inset plots the inverse susceptibility $\chi$$^{-1}$ versus temperature for $H \parallel c$ (up panel) and $H \parallel ab$ (bottom panel),  respectively. The black lines represent  Curie-Weiss fits.} 
	\label{fig:3}
\end{figure}
%%%%%%%%%%%%%%%%%%%%%%%%%

%%%%%%%%%%%%%%%%%%%%%%%
\begin{figure}[!htp]
	\centering
	\includegraphics[width=0.44\textwidth]{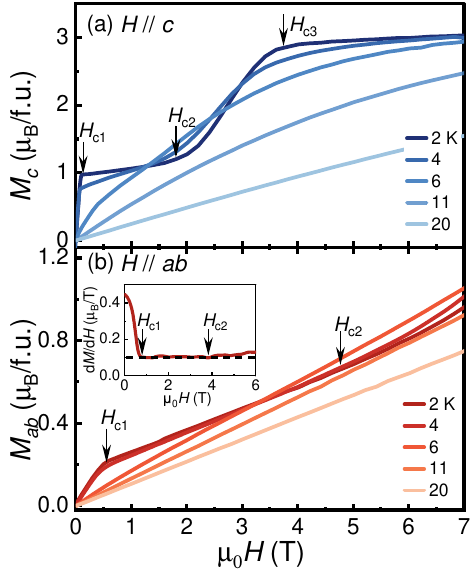}
	\caption{(Color online) Field-dependent magnetization $M(H)$ of NdAlGe %collected
		at various temperatures for $H \parallel c$ (a) and $H \parallel {ab}$ (b). The selected temperatures cover both the AFM/FIM and the PM states. The arrows mark the saturation field ($H_{c3}$) and the two critical fields ($H_{c1}$ and $H_{c2}$) where NdAlGe undergoes metamagnetic transitions. The inset in panel (b) shows the derivative of the in-plane magnetization $M_{ab}$ with respect to the magnetic field at 2\,K.}  
	\label{fig:4}
\end{figure}
%%%%%%%%%%%%%%%%%%%%%%

%%%%%%%%%%%%%%%%%%%%%%%%%%%%%%
\begin{figure*}[!htb]
	\centering
	\includegraphics[width=0.9\textwidth]{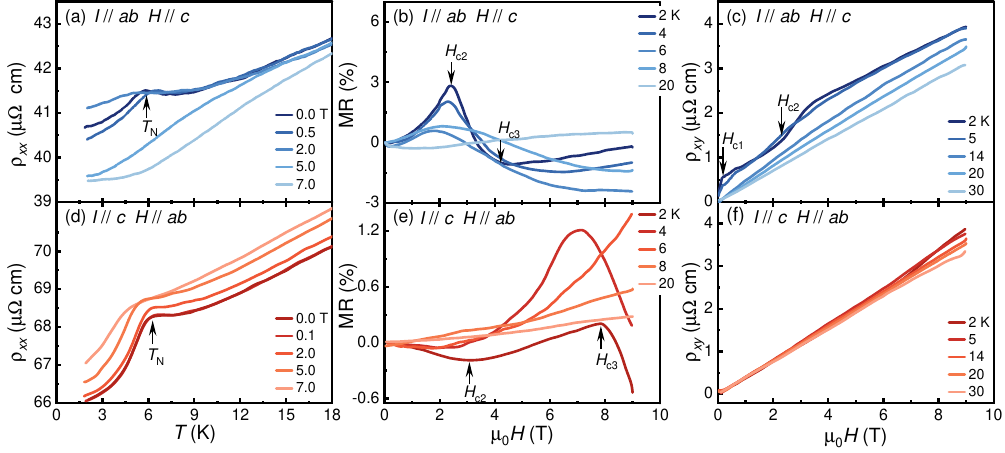}
	\caption{(Color online) (a) Temperature-dependent electrical resistivity $\rho_{xx}(T)$ for NdAlGe collected at various magnetic fields. (b) Field-dependent electrical resistivity $\rho_{xx}(H)$ collected at various temperatures. Here the $\rho_{xx}(H)$ is presented as magnetoresistivity, MR =  [$\rho_{xx}(H) -\rho_{xx}(0)]/\rho_{xx}(0)$, where $\rho_{xx}(0)$ is the zero-field electrical resistivity.	(c) Field-dependent Hall resistivity $\rho_{xy}(H)$ collected at various temperatures. The data in panels (a)-(c) were collected by applying the magnetic field along the $c$ axis ($H \parallel c$) and the electric current within the $ab$ plane ($I \parallel ab$). The analogous results for $H \parallel ab$ and $I \parallel c$ are shown in panels (d)-(f). The arrows in panels (a) and (d) mark the SDW transition at $T_\mathrm{N}$, while the arrows in panels (b), (c), and (e) mark the critical fields $H_{c1}$ and $H_{c2}$, as well as the saturation field $H_{c3}$.}
	\label{fig:5}
\end{figure*}
%%%%%%%%%%%%%%%%%%%%%%%%%%%%%%

The magnetic properties of the NdAlGe single crystal were
first characterized by magnetic susceptibility measurements, using both
field-cooling (FC) and zero-field-cooling (ZFC) protocols in applied
magnetic fields of 0.01 and 0.1\,T.
Two successive magnetic transitions, indicated by arrows
in Fig.~\ref{fig:2}(a), could be clearly identified in the $\chi(T)$ data recorded in a small magnetic field (0.01\,T)
applied along the $c$ axis or within the $ab$ plane.
Recent neutron diffraction studies
attribute the first magnetic transition, at $T_\mathrm{N} = 6.1$\,K, 
to the paramagnetic (PM) to incommensurate SDW transition~\cite{Yang2023,Dhital2023}. The second transition, at $T_\mathrm{FIM} = 4.5$\,K, is attributed to the 
subsequent commensurate FIM transition. Both transitions are clearly
reflected also in the heat capacity data~\cite{Dhital2023}. 
Between $T_\mathrm{FIM}$ and $T_\mathrm{N}$, the ZFC and
FC susceptibility data are practically identical, confirming the
antiferromagnetic (AFM) nature of the first transition. Below
$T_\mathrm{FIM}$, the ZFC and FC curves split due to the formation
of a FIM order. Upon increasing the magnetic field to
0.1\,T, both transitions at $T_\mathrm{N}$ and  $T_\mathrm{FIM}$ are
robust in the $\chi_{ab}(T)$ case, while $T_\mathrm{FIM}$
becomes almost invisible in the $\chi_c(T)$ case [see Fig.~\ref{fig:2}(b) and (d)]. 
This is due to the occurrence of a metamagnetic transition
for $H \parallel c$, with the FIM transition occurring only
below 0.1\,T (see further).
%\tcr{This type of behavior is attributed to the occurrence of metamagnetic transition near 0.1 T  so that the $T_\mathrm{FIM}$ can be only observed blow 0.1 T for $H \parallel c$.}
%suggests stronger magnetic interactions within the $ab$ plane than along the $c$ axis, consistent with the layered structure of NdAlGe, whose out-of-plane  lattice parameter is three times larger than that in plane.
%

At high temperatures, both $\chi_{ab}(T)$ and $\chi_c(T)$ show a
Curie-Weiss behavior [see $\chi(T)^{-1}$ in the inset of Fig.~\ref{fig:3}].
A Curie-Weiss fit yields effective moments {\textmu}$_\mathrm{eff} = 3.760(2)$
and 3.802(3)~{\textmu}$_\mathrm{B}$ and paramagnetic Curie temperatures
$\theta_\mathrm{p} = 8.34(7)$ and $-6.09(13)$\,K for $H \parallel c$
and $H \parallel ab$, respectively. Both effective moments are
close to the theoretical value for the free Nd$^{3+}$ ions 
($4f^3$, 3.62~{\textmu}$_\mathrm{B}$). The positive Curie temperature indicates
dominant ferromagnetic (FM) interactions along the $c$ axis,
whereas the negative value for fields lying in the $ab$ plane, suggests
dominant AFM interactions, a result consistent
with the neutron scattering data~\cite{Yang2023}. NdAlGe also exhibits
a strong out-of-plane magnetic anisotropy. As shown by the
$\chi_c$/$\chi_{ab}$ data in Fig.~\ref{fig:3}, the anisotropy
starts to increase below the onset of the AFM order, with
$\chi_c$/$\chi_{ab}$ reaching almost 70 at base temperature.

The out-of-plane magnetic anisotropy is also reflected in the
field-dependent magnetization measurements (see Fig.~\ref{fig:4}).
For example, for $H \parallel c$, the 2\,K magnetization reaches
a saturation value of 2.9\,{\textmu}$_\mathrm{B}$ at {\textmu}$_0H_{c3} = 3.8$\,T.
Such saturation magnetization is slightly smaller than the theoretical
value of 3.27\,{\textmu}$_\mathrm{B}$ expected for the
$J = 9/2$ Nd$^{3+}$ ions, yet consistent with previous studies~\cite{Yang2023}.
Conversely, for $H \parallel ab$, no saturation is observed
up to 7\,T, and the in-plane magnetization $M_{ab}$ is significantly
smaller than the out-of-plane magnetization $M_{c}$. Such robust magnetization against external in-plane magnetic fields is most likely attributed to the single-ion anisotropy induced by the crystalline electric-field effect. For $H \parallel c$, NdAlGe undergoes two metamagnetic transitions
below $T_\mathrm{N}$ as the field increases. For $H \parallel ab$,
the metamagnetic transitions are less evident, but still 
observable in the  derivative of magnetization with respect to
the magnetic field [see d$M$/d$H$ in the inset of Fig.~\ref{fig:4}(b)].
At base temperature, for $H \parallel c$, NdAlGe 
undergoes two metamagnetic
transitions, at {\textmu}$_0H_{c1} = 0.1$\,T and  {\textmu}$_0H_{c2} = 2.1$\,T,  
which become 0.6\,T and 3.6\,T, respectively, for $H \parallel ab$. In the latter case, also the saturation field {\textmu}$_0H_{c3}$ $\sim $
11.5\,T~\cite{Dhital2023} is markedly %significantly
larger than the value for $H \parallel c$ (3.8\,T).
Note that, all the metamagnetic transitions are very sensitive to
the growth conditions. Similar to our case, other NdAlGe
crystals grown by the flux method show distinct metamagnetic transitions~\cite{Yang2023,Zhao2022},
while these are absent in crystals grown by the floating-zone
method~\cite{Kikugawa2023}. Such diverse properties are most likely
attributed to the different level of Ge vacancies in NdAlGe single
crystals. Indeed, as discussed in detail for the isostructural
CeAlGe~\cite{Puphal2019}, crystals grown by the floating-zone method
are Al deficient, while those grown from molten flux are rich in Al. 
% It is not very clear how the Al-deficiency is related to the Ge vacancies!

%%%%%%%%%%%%%%%%%%%%%%%%%%%%
\begin{figure}[!htp]
	\centering
	\includegraphics[width=0.46\textwidth]{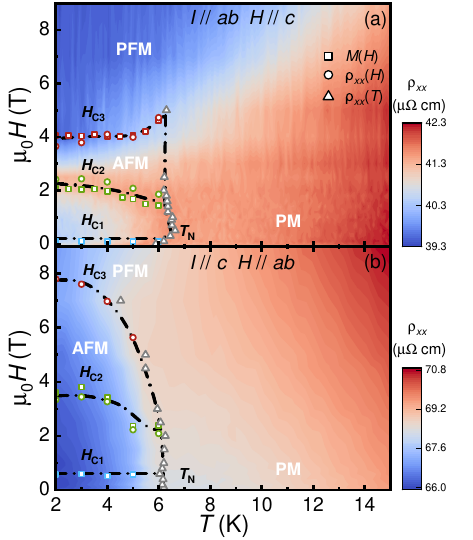}
	\caption{(Color online) Magnetic phase diagram of a NdAlGe
		single crystal with the magnetic field applied along the $c$
		axis (a) and within the $ab$ plane (b). The critical fields were
		determined from $M(H)$ and $\rho_{xx}(H)$ data in Fig.~\ref{fig:4}
		and Fig.~\ref{fig:5}. The magnetic transition temperatures were
		determined from the $\rho_{xx}(T)$ data shown in Fig.~\ref{fig:5}.
		The background color represents the magnitude of $\rho_{xx}(T,H)$.
		The PM, FIM, and PFM denote the paramagnetic, fe\-rri\-mag\-ne\-tic,
		and polarized ferromagnetic states, respectively. Since the in\-commensurate
		SDW phase is limited to a narrow temperature and field range, it is not marked in
		the phase diagram. The dash-dotted lines are guides to the eyes.}
	\label{fig:6}
\end{figure}
%%%%%%%%%%%%%%%%%%%%%%%%%%%%%

We also performed systematic temperature and field dependence of the electrical
resistivity on NdAlGe. As indicated by the arrows in
Fig.~\ref{fig:5}(a) and (d), the SDW transition at $T_\mathrm{N} \sim 6$\,K
is clearly visible in the $\rho_{xx}(T)$ data. Yet, no distinct anomaly was observed at $T_\mathrm{FIM}$, implying
that the modulation due to the FIM magnetic structure has
negligible effects on the magnetic scattering of electrons. Upon
increasing the magnetic field to 2\,T, the magnetic transition becomes
less visible for $H \parallel c$, while $T_\mathrm{N}$ is almost
independent of the applied field.
For {\textmu}$_0H > 2$\,T, the transition temperature is best
tracked by the derivative of resistivity with respect to
temperature. By contrast, for $H \parallel ab$, the magnetic
transition is always visible, but it moves slowly to lower
temperatures. Figures~\ref{fig:5}(b) and (e) show the field-dependent
electrical resistivity $\rho_{xx}(H)$ at selected temperatures, here
presented as magnetoresistivity (MR). Below $T_\mathrm{N}$, the MR exhibits clear field-induced transitions.
As indicated by the arrows and compared with the $M(H)$ data
in Fig.~\ref{fig:4}, the first transition corresponds to the
metamagnetic transition at $H_{c2}$, while the second one is consistent
with the saturation of magnetization at $H_{c3}$. For both orientations,
the MR is rather small, reaching at most 3\% 
at $H_{c2}$ for $H \parallel c$
[see Fig.~\ref{fig:5}(b)]. Note that, the first metamagnetic
transition at $H_{c1}$ is absent in the MR data but, 
below $T_\mathrm{N}$ and for $H \parallel c$, 
the Hall resistivity $\rho_{xy}(H)$
exhibits clear anomalies at both $H_{c1}$ and $H_{c2}$ [see Fig.~\ref{fig:5}(c)].
While for $H \parallel ab$, although both the $M(H)$ and $\rho_{xx}(H)$ data exhibit metamagnetic transitions, no distinct anomaly can be found in $\rho_{xy}(H)$ [see Fig.~\ref{fig:5}(f)]. Such different field responses suggest that the anomalies in $\rho_{xy}(H)$ for $H \parallel c$ are most likely attributed to the field induced topological spin textures in NdAlGe.
Furthermore, the positive slope of $\rho_{xy}(H)$ indicates 
that hole carriers are dominant in NdAlGe single crystal,
consistent with the band-structure calculations~\cite{Yang2023}.

%%%%%%%%%%%%%%%%%%%%%%%%%%%%%%%
\begin{figure}[!htp]
	\centering
	\includegraphics[width=0.46\textwidth]{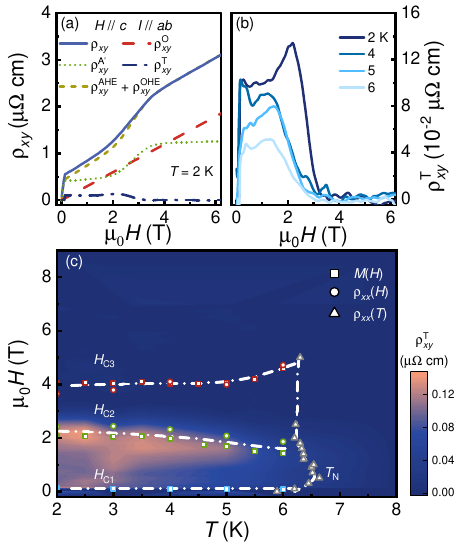}
	\vspace{-2mm}
	\caption{(Color online) (a) Analysis of the Hall resistivity for $H \parallel c$ at $T$ = 2\,K. The other $\rho_{xy}(H)$ data were analyzed using the same procedure. Solid, dashed, dotted, and short-dashed lines represent the measured total Hall resistivity $\rho_{xy}$, the ordinary Hall resistivity $\rho_{xy}^\mathrm{O}$, the anomalous Hall resistivity $\rho_{xy}^\mathrm{A'}$, and the sum of $\rho_{xy}^\mathrm{A'}$ and $\rho_{xy}^\mathrm{O}$, respectively. The extracted topological Hall resistivity $\rho_{xy}^\mathrm{T}$ is shown by the dash-dotted line. (b)  Extracted field-dependent topological Hall resistivity $\rho_{xy}^\mathrm{T}(H)$ at different temperatures below $T_\mathrm{N}$.
		(c) Magnetic phase diagram of a NdAlGe single crystal with the field applied along the $c$ axis [as in Fig.~\ref{fig:6}(a)]. The background color represents the magnitude of $\rho_{xy}^\mathrm{T}(H)$ at various temperatures.} 
	\label{fig:7}
\end{figure}
%%%%%%%%%%%%%%%%%%%%%%%%%%%%%%

Figure~\ref{fig:6} summarizes the magnetic phase diagram of NdAlGe for both $H \parallel c$ and $H \parallel ab$. In general, the magnetic phase diagrams share some common features. For $H \parallel c$, the SDW transition temperature $T_\mathrm{N}$ 
increases slightly for magnetic fields up to 1\,T, 
to become almost temperature independent for fields up to 3\,T.
While, for $H \parallel ab$, $T_\mathrm{N}$ is smoothly suppressed to
lower temperatures, reaching $\sim 4.5$\,K at {\textmu}$_0H = 7$\,T.
Upon further increasing the magnetic field, Nd moments become
fully polarized for {\textmu}$_0H \gtrsim 4$--5\,T and
{\textmu}$_0H \gtrsim 8$\,T for $H \parallel c$ and $H \parallel ab$,
respectively.
%\tcr{\sout{This is consistent with the stronger  magnetic coupling within the $ab$ plane.}}
In the SDW/FIM state, by applying a 
magnetic field, Nd moments undergo two successive  
metamagnetic transitions at $H_{c1}$ and $H_{c2}$ before entering the polarized ferromagnetic (PFM) state at $H > H_{c3}$. At base temperature, the in-plane saturation field $H_{c3}$ is almost twice larger than that out-of-plane. 
While, for $H \parallel ab$, $H_{c3}$ decreases as the temperature approaches the magnetic order at $T_\mathrm{N}$, it shows an opposite behavior for $H \parallel c$, consistent with previous studies~\cite{Zhao2022}. As a consequence, close to $T_\mathrm{N}$, 
the saturation fields along both directions are comparable, implying a
strong competition between the in-plane and out-of-plane magnetic interactions.
According to previous neutron scattering results~\cite{Yang2023}, NdAlGe undergoes a magnetic transition from SDW to a FIM state at temperatures just below $T_\mathrm{N}$. However, such transition might be easily destroyed by an applied magnetic field larger than $H_{c1}$, 
thus demanding further studies. %investigations.

%%%%%%%%%%%%%%%%%%%%%%%%%%%%%%%%
\begin{figure*}[htbp]
	\centering
	\includegraphics[width=0.9\textwidth]{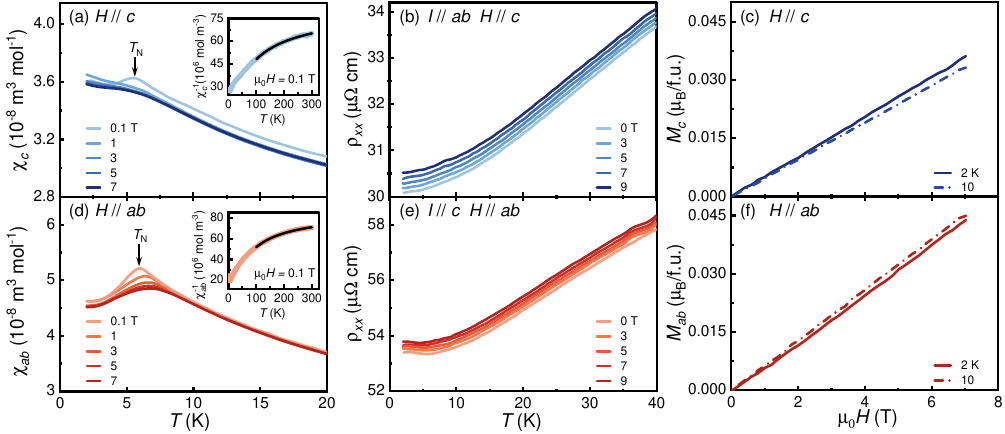}
	\caption{(Color online) 
		Temperature-dependent FC magnetic susceptibility $\chi_c(T)$ (a) 
		and electrical resistivity $\rho_{xx}(T)$ (b) collected under
		various magnetic fields, and field-dependent magnetization $M_c(H)$
		(c) collected at 2 and 10\,K for a SmAlGe single crystal.
		The magnetic field was applied along the $c$ axis, while the current was applied within the $ab$ plane.
		The analogous results for $H \parallel ab$ and $I \parallel c$ are shown in panels (d)-(f). The insets in panels (a) and (b) show the 0.1\,T inverse susceptibility $\chi_c^{-1}$ and $\chi_{ab}^{-1}$ versus temperature respectively, with the black lines representing Curie-Weiss fits. 
		The arrows in panels (a) and (d) mark the AFM transition at $T_\mathrm{N}$.
		Note that, unlike in NdAlGe, the ZFC and FC susceptibilities of SmAlGe are almost identical.}
	\label{fig:8}
\end{figure*}
%%%%%%%%%%%%%%%%%%%%%%%%%

Considering that near the critical fields the Hall resistivity data for $H \parallel c$ exhibit clear anomalies, 
most likely related to the topological spin textures in NdAlGe, we further analyzed the Hall resistivity. 
In general, the measured Hall resistivity in a magnetic material can be written as $\rho_{xy}$ = $\rho_{xy}^\mathrm{O}$ + $\rho_{xy}^\mathrm{A}$, where $\rho_{xy}^\mathrm{O}$ and $\rho_{xy}^\mathrm{A}$ represent the ordinary and anomalous Hall resistivity, respectively. The second term can be further split 
into a conventional anomalous Hall resistivity $\rho_{xy}^\mathrm{A'}$ and a topological Hall resistivity $\rho_{xy}^\mathrm{T}$. While $\rho_{xy}^\mathrm{A'}$ is mostly determined by the momentum-space Berry curvatures, 
$\rho_{xy}^\mathrm{T}$ is attributed to the topological spin textures with a finite scalar 
spin chirality (also known as real-space Berry curvatures), e.g., magnetic skyrmions~\cite{Neubauer2009,Kanazawa2011,Kurumaji2019,Schulz2012,Qin2019,Shang2021,roychowdhury2023,khanh2020,zhang2023-1,yu2012}. In Fig.~\ref{fig:7}(a), we show the different contributions
to the Hall resistivity at 2\,K versus magnetic field, with the
other temperatures showing similar results. 
Here, we use $M(H)$ to obtain $\rho_{xy}^\mathrm{A'}(H)$ from 
$\rho_{xy}^\mathrm{A'}(H) = R_{s} M(H)$, where $R_{s}$ is a
field-independent scaling factor. We note that the use of $\rho_{xx}^2M(H)$ or $\rho_{xx}M(H)$ leads to comparable $\rho_{xy}^\mathrm{T}$ values, since $\rho_{xy} \ll \rho_{xx}$. After subtracting the $\rho_{xy}^\mathrm{O}(H)$ and $\rho_{xy}^\mathrm{A'}(H)$ components from the measured $\rho_{xy}(H)$, the extracted topological contribution $\rho_{xy}^\mathrm{T}(H)$ is shown by a dash-dotted line in Fig.~\ref{fig:7}(a). The $\rho_{xy}^\mathrm{T}(H)$ at various temperatures are shown as a background in Fig.~\ref{fig:7}(b).
Obviously, in the field range between $H_{c1}$ and $H_{c2}$, where Nd$^{3+}$ moments undergo a first and a second me\-ta\-mag\-net\-ic transition, $\rho_{xy}^\mathrm{T}(H)$ is particular evident, to become almost invisible at temperatures above 6\,K. Such field and temperature dependence of $\rho_{xy}^\mathrm{T}(H,T)$ is reminiscent of possible topological spin textures in NdAlGe~\cite{Neubauer2009,Kanazawa2011,Kurumaji2019,Schulz2012,Qin2019,Shang2021,roychowdhury2023}.

%%%%%%%%%%%%%%%%%%%%%%%%%
\subsection{Magnetic and transport properties of SmAlGe}\label{sec:SAG}
%%%%%%%%%%%%%%%%%%%%%%%%%

%%%%%%%%%%%%%%%%%%%%%%%%%%%%%%
\begin{figure}[!htp]
	\centering
	\includegraphics[width=0.45\textwidth]{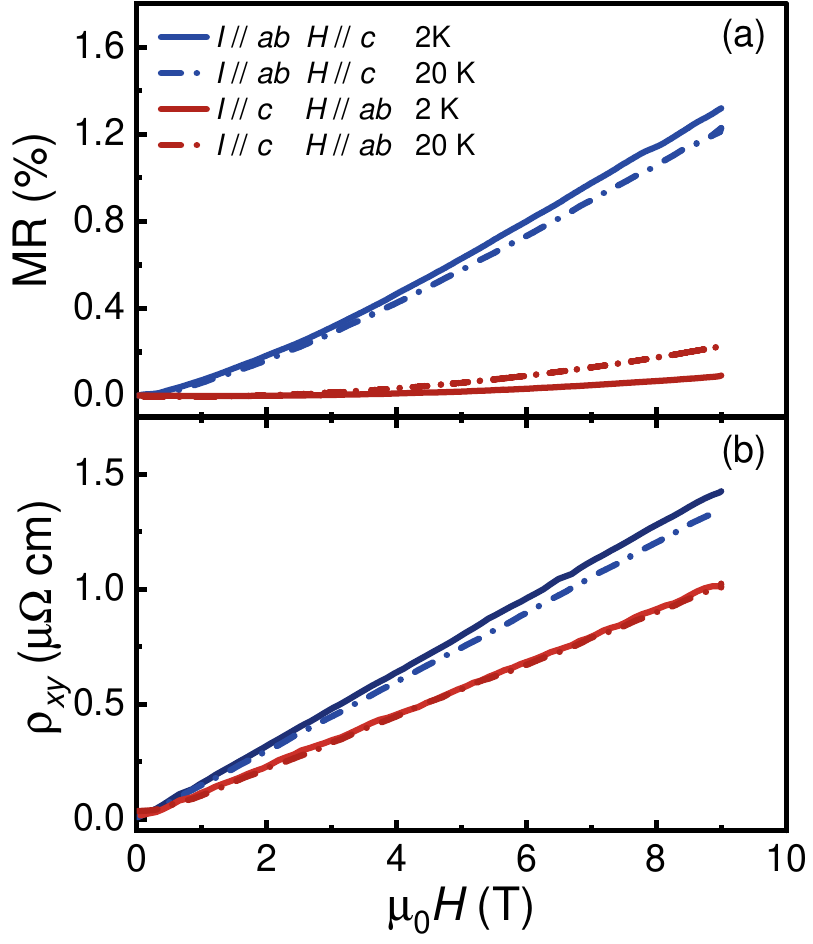}
	\caption{(Color online) Field-dependent magnetoresistivity (a)
		and Hall resistivity (b) collected in the AFM (2\,K) and
		PM states (20\,K) of a SmAlGe single crystal for
		$H \parallel c$ and $H \parallel ab$, respectively.} 
	\label{fig:9}
\end{figure}
%%%%%%%%%%%%%%%%%%%%%
Different from the NdAlGe single crystal, the other $f$-series based compound SmAlGe exhibits trivial magnetic and transport properties. For $H \parallel c$, as indicated by the arrow in Fig.~\ref{fig:8}(a), the $\chi_c(T)$ shows a peak-like anomaly at $T_\mathrm{N}$ = 5.6\,K in a field of 0.1\,T, 
typical of an AFM transition, as further confirmed by the linear field-dependent magnetization $M(H)$ [see Fig.~\ref{fig:8}(c) and (f)]. Upon increasing the magnetic field above 1\,T, the AFM peak evolves into a broad shoulder, 
while $T_\mathrm{N}$ increases slightly to reach $\sim 7$\,K at {\textmu}$_0H = 7$\,T. For $H \parallel ab$, 
the peak in $\chi_{ab}(T)$ is also robust against the applied field, with $T_\mathrm{N}$ increasing from 5.9\,K at 0.1\,T to 6.7\,K at 7\,T [see Fig.~\ref{fig:8}(d)].
At high temperatures, both $\chi_c(T)$ and $\chi_{ab}(T)$ show a Curie-Weiss behavior with a relatively large $\chi_0$ contribution compared to NdAlGe [see, e.g., $\chi_c(T)^{-1}$ in the inset of Fig.~\ref{fig:8}(a)]. The Curie-Weiss fit (see solid line) yields effective moments {\textmu}$_\mathrm{eff} = 0.75(2)$ and 0.68(1)~{\textmu}$_\mathrm{B}$ and paramagnetic Curie temperatures $\theta_\mathrm{p} = -8(3)$ and 6(2)\,K for $H \parallel c$ and $H \parallel ab$, respectively. The derived effective moments are close to the theoretical low-temperature value 
% Note: Sm3+ has mu_exp = 1.79 muB at RT, due to the closeness of the excited states.
% Hence, it is a pure coincidence that we measure at low T, where kB T << Delta E.
of free Sm$^{3+}$ ions ($4f^5$, 0.85~{\textmu}$_\mathrm{B}$). Unlike the NdAlGe case,
in SmAlGe, the AFM interactions are dominant along the $c$ axis,
while FM interactions dominate in the $ab$ plane. Moreover, SmAlGe is magnetically isotropic, here reflected in the almost identical magnetization along both directions.

%%%%%%%%%%%%%%%%%%%%%%%%
\begin{figure}[!htp]
	\centering
	\includegraphics[width=0.48\textwidth]{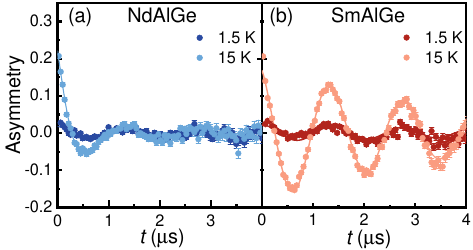}
	\caption{(Color online) Time-domain wTF-{\textmu}SR spectra collected in a transverse muon-spin configuration ($\boldsymbol{P}_\mu$ $\perp$ $\boldsymbol{S}_\mu$) in NdAlGe (a) and SmAlGe (b) single crystals. 
		Here, representative spectra collected at 1.5\,K (FIM/AFM state) and at 15\,K (PM state) in a weak transverse field of 5\,mT are shown. The solid lines through the data are fits to Eq.~\eqref{eq:1}.} 
	\label{fig:10}
\end{figure}
%%%%%%%%%%%%%%%%%%%%%%%

We also measured the temperature-dependent in-plane and out-of-plane electrical resistivity $\rho_{xx}(T)$ under various magnetic fields up to 9\,T. As shown in Fig.~\ref{fig:8} (b) and (e), no clear 
anomalies can be tracked near the AFM transition in $\rho_{xx}(T)$. All the $\rho_{xx}(T)$ curves share the same temperature dependence, implying the absence of metamagnetic transition in SmAlGe,
consistent with the $M(H)$ data in Fig.~\ref{fig:8}. As shown in Fig.~\ref{fig:9}(a), SmAlGe exhibits positive MR both in the AFM and the PM states, with the highest MR value reaching $\sim 1.3$\% for $H \parallel c$ at 9\,T, %$\mu_0H$ = 9\,T, 
slightly smaller than that of NdAlGe (see Fig.~\ref{fig:5}). 
In both SmAlGe and NdAlGe, the in-plane MR ($I \parallel ab$) is larger
than the out-of-plane MR ($I \parallel c$).
For both field directions, the 
MR curve at 20\,K is almost identical to
that at 2\,K, indicating that the magnetic order has negligible effects
on the magnetotransport properties of SmAlGe. 
As a consequence, as shown in Fig.~\ref{fig:9} (b), the Hall resistivity is almost linear in field, with the hole carriers being again dominant.

%%%%%%%%%%%%%%%%%
\subsection{Muon-spin rotation and relaxation}\label{sec:muSR}

%%%%%%%%%%%%%%%%%
\begin{figure}[!htp]
	\centering
	\includegraphics[width=0.45\textwidth]{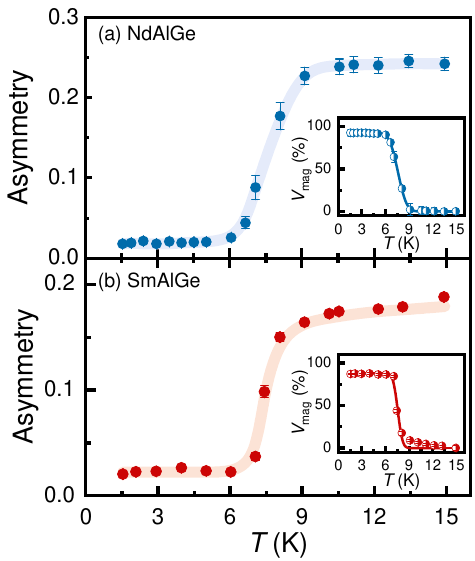}
	\caption{(Color online) Temperature dependence of the asymmetry $A_\mathrm{NM}$
		obtained from the wTF-{\textmu}SR spectra of NdAlGe (a)
		and SmAlGe (b) single crystals. The insets show the magnetic volume fraction vs temperature. Here, lines are fits to the phenomenological function given in Eq.~\eqref{eq:2}.} 
	\label{fig:11}
\end{figure}
%%%%%%%%%%%%%%%%%%%%%%%
To investigate the magnetic properties of $RE$AlGe ($RE$ = Nd, Sm) at a microscopic level, we performed a series of {\textmu}SR measurements at temperatures covering both their FIM/AFM and PM states. 
The magnetic transition temperature $T_\mathrm{N}$ and the magnetic volume fraction of NdAlGe and SmAlGe single crystals were established by means of wTF-{\textmu}SR measurements. 
As shown in Fig.~\ref{fig:10}, a weak transverse field of 5\,mT was applied perpendicular to the initial muon-spin direction in the PM state (e.g., 15\,K), where it leads to coherent oscillations. 
In the FIM/AFM state (e.g., 1.5\,K), the internal fields are
much larger than the applied 5\,mT field (see ZF-{\textmu}SR below). Therefore, muon spins precess with
frequencies that reflect the internal fields at the muon-stopping sites rather than the weak applied field. Considering that the magnetic order normally leads to a very fast
muon-spin relaxation in the first tenths of a microsecond, the wTF-{\textmu}SR spectra can be analyzed using the function:
%%%%%%%%%%%%%%%%
\begin{equation}
	\centering
	A_\mathrm{wTF}(t) = A_\mathrm{NM}\cos(\omega t + \phi) e^{-\lambda t},  
	\label{eq:1}
\end{equation}
%%%%%%%%%%%%%%%%%
where $A_\mathrm{NM}$ is the initial muon-spin asymmetry (i.e., the amplitude of the oscillation) for muons implanted in the  
nonmagnetic (NM) or PM fraction of the NdAlGe and SmAlGe
single crystals; $\omega$ ($\equiv \gamma_{\mu}$$B^\mathrm{int}$) is the muon-spin precession frequency, with $\gamma_{\mu}$ = 2$\pi$ $\times$ 135.5\,MHz/T the muon gyromagnetic ratio and $B^\mathrm{int}$ the local field sensed by the implanted muons; $\phi$ is the initial phase, and $\lambda$ is the muon-spin relaxation rate. Note
that, in the magnetically ordered state, the initial
fast {\textmu}SR relaxation was discarded and only the long-time slow-relaxing asymmetry was analyzed
(see the 1.5\,K dataset in Fig.~\ref{fig:10}).
The resulting signal is only due to the nonmagnetic sample fraction, as
confirmed by a $B^\mathrm{int}$ value almost identical to the applied magnetic field (here, $\sim$ 5\,mT).

%%%%%%%%%%%%%%
\begin{figure}[!htp]
	\centering
	\includegraphics[width=0.48\textwidth]{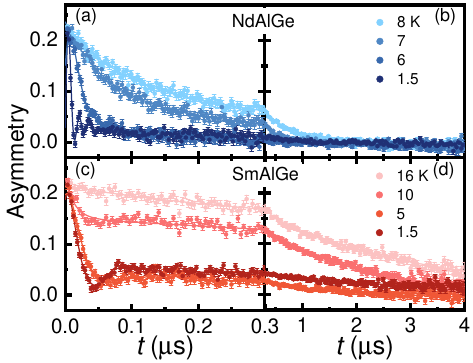}
	\caption{(Color online) Representative ZF-{\textmu}SR spectra collected in a transverse muon-spin configuration ($\boldsymbol{P}_\mu$ $\perp$ $\boldsymbol{S}_\mu$) at temperatures covering both the PM and FIM/AFM states of a NdAlGe single crystal in a short (a) and a long time window (b). The analogous results for SmAlGe are shown in panels (c) and (d). The short-time spectra in panels (a) and (c) illustrate the coherent oscillations caused by the long-range magnetic order. Solid lines through the data are fits to Eq.~\eqref{eq:3} (FIM/AFM state) and Eq.~\eqref{eq:4} (PM state).} 
	\label{fig:12}
\end{figure}
%%%%%%%%%%%%%%%%%%%%%

The derived wTF-{\textmu}SR asymmetry values $A_\mathrm{NM}$ are summarized in Fig.~\ref{fig:11} for NdAlGe and SmAlGe single crystals. In the PM state, all the implanted muons precess at the same frequency $\omega$. As the temperature approaches the 
magnetic transition, only the muons implanted in the remaining PM/NM 
phase still precess at the frequency $\omega$, here reflected in a progressive reduction of the asymmetry.
Below $T_\mathrm{N}$, most of the muons precess at frequencies that are determined by the long-range magnetic order in both crystals. The PM (or NM) sample fraction is then determined from the oscillation amplitude. In both NdAlGe and SmAlGe, $A_\mathrm{NM}$ starts to decrease near the onset of magnetic order. Note that, in the SmAlGe case, $A_\mathrm{NM}$ is affected by the strong spin fluctuations in the PM state (see ZF-{\textmu}SR below). As a consequence, its $A_\mathrm{NM}(T)$ shows a weak temperature dependence even at $T > T_\mathrm{N}$. 
The temperature evolution of the magnetic volume fraction can be estimated
from $V_\mathrm{mag}(T) = 1 - A_\mathrm{NM}(T)/A_\mathrm{NM} (T > T_\mathrm{N}$).
The obtained $V_\mathrm{mag}(T)$ are summarized in the insets of Fig.~\ref{fig:11}
and fitted using the phenomenological function:
%%%%%%%%%%%%%%%%%%
\begin{equation}
	V_\mathrm{mag}(T) = V_\mathrm{mag}(0) \frac{1}{2}\left[1-\mathrm{erf}\left(\frac{T - T_\mathrm{N}}{\sqrt{2} \Delta T}\right)\right].
	\label{eq:2}
\end{equation}
%%%%%%%%%%%%%%%%%%%%
Here, $V_\mathrm{mag}(0)$, $\Delta$$T$, and erf($T$) are the
zero-temperature magnetic volume fraction, the magnetic transition width, and the 
error function, respectively. 
As shown by solid lines in the insets of Fig.~\ref{fig:11}, for NdAlGe, we obtained $V_\mathrm{mag}(0)$ = 92.5(4)\%,
$\Delta$$T$ = 0.86(3)\,K, and $T_\mathrm{N}$ = 7.58(2)\,K.
While for SmAlGe, $V_\mathrm{mag}(0)$ = 87(2)\%, $\Delta$$T$ = 0.48(9)\,K, and $T_\mathrm{N}$ = 7.56(6)\,K. 
Clearly, at low temperatures, both crystals can be considered as fully magnetically ordered. 
Note that, the transition tem\-per\-a\-tures $T_\mathrm{N}$ determined from $V_\mathrm{mag}(T)$  are slightly higher 
%fully magnetically ordered. Note that, the transition tem\-per\-a\-tures
than the values from the magnetic- or transport data. However, the offsets at $\sim$6.4\,K and $\sim$6.8\,K for NdAlGe and SmAlGe are in very good agreement with the magnetic and transport data.

The large gyromagnetic ratio of muons, combined with their availability as 100\% spin-polarized beams, makes ZF-{\textmu}SR a very sensitive probe for investigating magnetic materials~\cite{Yaouanc2011,Hillier2022}. To study the local magnetic properties of NdAlGe and SmAlGe single crystals, we collected a series of ZF-{\textmu}SR spectra at temperatures covering both the PM and FIM/AFM states. The ZF-{\textmu}SR spectra in the magnetically ordered state ($T < T_\mathrm{N}$) are characterized by highly damped oscillations, typical of a long-range magnetic order with complex magnetic structures [see Fig.~\ref{fig:12}(a) and (c)], superimposed on a slowly decaying relaxation, observable only at long times [see Fig.~\ref{fig:12}(b) and (d)]. 
Therefore, the ZF-{\textmu}SR spectra of NdAlGe and SmAlGe were analyzed using the following model: 
%%%%%%%%%%%%%%%%%%%%%%
\begin{equation}
	A_\mathrm{ZF}(t) = \sum_{i=1}^{2} A_i \left[{\alpha}\cos(\omega_i t + \phi) e^{-\lambda^\mathrm{T}_i t} + (1 - \alpha) e^{-\lambda^\mathrm{L}_i t}\right].
	\label{eq:3}
\end{equation}
%%%%%%%%%%%%%%%%%%%%%%
Here, $\alpha$ and $(1 - \alpha)$ are the oscillating (i.e., transverse) and 
nonoscillating (i.e., longitudinal) fractions of the {\textmu}SR signal, respectively, whose initial total asymmetry is equal to $A_i$; $\lambda^\mathrm{T}_i$ 
and $\lambda^\mathrm{L}_i$ represent the transverse and longitudinal relaxation rates, with the former reflecting the intrinsic internal field distribution and the latter related solely to the spin fluctuations;
$\omega_i$ ($\equiv \gamma_{\mu}$$B^\mathrm{int}_i$) and $\phi$ being the same as in Eq.~\eqref{eq:1}. In the PM state ($T > T_\mathrm{N}$), the oscillations disappear, but the {\textmu}SR spectra 
still exhibit a rather fast muon-spin depolarization, implying the presence of strong spin fluctuations, here further confirmed by LF-{\textmu}SR measurements (see below).
The ZF-{\textmu}SR spectra in the PM state were then modeled by:
%\setlength{\belowdisplayskip}{0pt} \setlength{\belowdisplayshortskip}{12pt}
%%%%%%%%%%%%%%%%%%%%%%
\begin{equation}
	A_\mathrm{ZF}(t) = \sum_{i=1}^{2} A_i  e^{-\lambda^\mathrm{L}_i t}.
	\label{eq:4}
\end{equation}
%%%%%%%%%%%%%%%%%%%%%%	
In both NdAlGe and SmAlGe single crystals, we found two nonequivalent muon-stopping sites, i.e., $i$ = 2. Similar expressions have been frequently used to analyze the {\textmu}SR data in other rare-earth based magnetic materials~\cite{Tran2018,Zhu2022}.

%%%%%%%%%%%%%%%%%%%%%%
\begin{figure*}[!htp] %[H]
	\centering
	\includegraphics[width=0.8\textwidth]{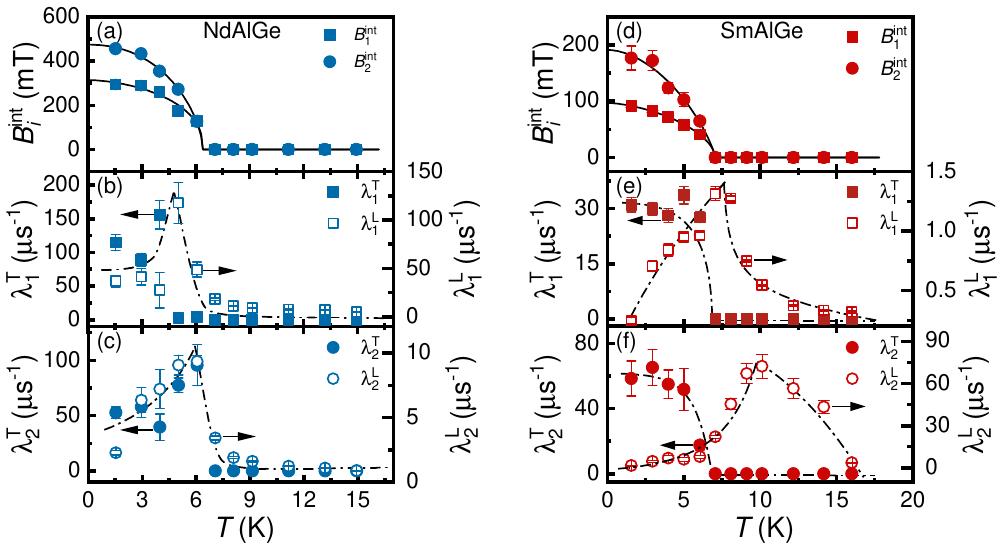}%\hspace{8mm}
	\caption{(Color online) (a) Temperature dependence of the internal field $B^\mathrm{int}_i(T)$ for NdAlGe. Temperature-dependent transverse $\lambda^\mathrm{T}_i(T)$ and longitudinal $\lambda^\mathrm{L}_i(T)$ muon-spin relaxation rates for site 1 (b) and site 2 (c) in NdAlGe. The analogus results for SmAlGe are shown in the panels (d)-(f). 
		Solid lines in panels (a) and (d) are fits to the equation described in the text; dashed-dotted lines in panels (b)-(c) and (e)-(f) are guides to the eyes.}
	\label{fig:13}
\end{figure*}
%%%%%%%%%%%%%%%%%%%%%%

The derived fit parameters from the ZF-{\textmu}SR data are summarized in Fig.~\ref{fig:13} for the NdAlGe (left) and SmAlGe single crystals (right), respectively. As can be seen in the top panels, both crystals show similar temperature-dependent $B^\mathrm{int}_i(T)$, resembling the typical mean-field type curve below $T_\mathrm{N}$. In both compounds, $B^\mathrm{int}_i(T)$ can be modeled by a phenomenological model $B^\mathrm{int}_i(T) = B^\mathrm{int}_i(0) \left[ 1 - (\frac{T}{T_\mathrm{N}})^{\gamma_i}\right]^{\delta_i}$, where $B^\mathrm{int}_i(0)$ is the internal magnetic field at zero temperature, while $\gamma_i$ and $\delta_i$ are two empirical parameters. As shown by solid lines in Fig.~\ref{fig:13}(a) and (d), the above model describes the data very well, yielding the parameters listed in Table~\ref{tab:1}. For
NdAlGe, the derived $B^\mathrm{int}_i(0)$ are 314(24) and 474(9)\,mT. While for SmAlGe, they are 97(1) and 191(12)\,mT.
Considering that NdAlGe and SmAlGe adopt the same LaPtSi-type crystal struc\-ture, the muon-stopping sites are expected to be similar in both cases.  Therefore, the larger internal fields in NdAlGe reflect the larger magnetic moment of Nd$^{3+}$ vs Sm$^{3+}$ ions.

The temperature dependence of the transverse and longitudinal {\textmu}SR relaxation rates $\lambda^\mathrm{T}(T)$ and $\lambda^\mathrm{L}(T)$ is
summarized in Fig.~\ref{fig:13}(b)-(c) for NdAlGe and in Fig.~\ref{fig:13}(e)-(f) for SmAlGe, respectively. In the PM state, due to the lack of a long-range order,
$\lambda^\mathrm{T}(T)$ is zero in both crystals. 
In the magnetically ordered state, $\lambda^\mathrm{T}(T)$
exhibits a clearly different temperature behavior. 
In NdAlGe, $\lambda^\mathrm{T}_1$ starts to increase below $T_\mathrm{N}$ and exhibits a significant jump at $T_\mathrm{FIM}$ $\sim$ 5\,K. On further lowering the temperature, 
$\lambda^\mathrm{T}_1$ decreases to $\sim$110\,{\textmu}s$^{-1}$ at 1.5\,K.
As for the second muon-stopping site, $\lambda^\mathrm{T}_2$
decreases continuously upon lowering the temperature, reaching $\sim 50$\,{\textmu}s$^{-1}$ at 1.5\,K. In contrast to the $\lambda^\mathrm{T}_1$, there is a clear drop at $T_\mathrm{FIM}$ in the $\lambda^\mathrm{T}_2$ channel. 
Note, however, that $B^\mathrm{int}_i(T)$
does not show any anomaly at $T_\mathrm{FIM}$,  
thus, suggesting that the modifications of magnetic structure are
too tiny to produce a measurable effect on the internal fields.
A similar behavior is encountered also in EuAl$_4$, where only significant changes to magnetic structure
lead to a clear anomaly in $B^\mathrm{int}(T)$~\cite{Zhu2022}.
Nonetheless, the AFM-FIM (or SDW-FIM) transition has a dramatic impact
on the field distribution, here reflected in a significant
jump/drop in $\lambda^\mathrm{T}_i$. Conversely, in SmAlGe, 
$\lambda^\mathrm{T}(T)$ exhibits a similar temperature dependence to
$B^\mathrm{int}(T)$. As shown in Fig.~\ref{fig:13}(e)-(f),
here $\lambda^\mathrm{T}$ becomes increasingly prominent as the temperature decreases below $T_\mathrm{N}$, reaching $\sim$30\,{\textmu}s$^{-1}$ and $\sim$60\,{\textmu}s$^{-1}$ at 1.5\,K for $\lambda^\mathrm{T}_1$ and $\lambda^\mathrm{T}_2$, respectively.
In either NdAlGe or SmAlGe, a very large $\lambda^\mathrm{T}$
at temperatures well below $T_\mathrm{N}$
is unusual for antiferromagnets and it implies a remarkably
inhomogeneous distribution of local fields in their magnetically ordered
state. The estimated half widths at half maximum (HWHM) of the
field distribution are listed in Table~\ref{tab:1}. Such enhanced
local-field distribution most likely arises from the complex
spatial arrangement of the rare-earth magnetic moments in SmAlGe
and NdAlGe, where a helical magnetic order has been detected by
neutron scattering~\cite{Yang2023}. 

The temperature-dependent longitudinal {\textmu}SR relaxation rates
$\lambda^\mathrm{L}(T)$ are also summarized in Fig.~\ref{fig:13}(b)-(c)
and Fig.~\ref{fig:13}(e)-(f). In NdAlGe, $\lambda^\mathrm{L}(T)$ exhibits a similar temperature dependence for both muon-stopping sites, typical of materials with a long-range magnetic order. $\lambda^\mathrm{L}_1(T)$ and $\lambda^\mathrm{L}_2(T)$ diverge near $T_\mathrm{FIM}$ and $T_\mathrm{N}$, respectively, followed by a significant decrease at $T < T_\mathrm{FIM}$ or $T_\mathrm{N}$,
confirming that spin fluctuations are the strongest close to the onset
of a magnetic order. More interestingly, the different temperatures where $\lambda^\mathrm{L}_i(T)$ diverges suggests that the first muon-stopping site is much closer to the Nd$^{3+}$ magnetic moments, hence, most relevant to the AFM-FIM transition in NdAlGe. In the studied temperature range, $\lambda^\mathrm{L}_1$ is always significantly larger than $\lambda^\mathrm{L}_2$. For example, at $T \sim$13\,K, $\lambda^\mathrm{L}_1$ $\sim$ 8\,{\textmu}s$^{-1}$ and $\lambda^\mathrm{L}_2$ $\sim$ 1\,{\textmu}s$^{-1}$; while at $T \sim$1.5\,K, $\lambda^\mathrm{L}_1$ $\sim$ 37\,{\textmu}s$^{-1}$ and $\lambda^\mathrm{L}_2$ $\sim$ 2.3\,{\textmu}s$^{-1}$. In NdAlGe, at $T \ll T_\mathrm{N}$, $\lambda^\mathrm{L}$ is almost 5 times larger than in the PM state, suggesting that the strong spin fluctuations persist in the magnetically ordered state. In SmAlGe, $\lambda^\mathrm{L}_1(T)$ exhibits a similar temperature dependence as in NdAlGe, with a divergence near the onset of the AFM order [see Fig.~\ref{fig:13}(e)]. Considering that $\lambda^\mathrm{L}_2(T)$
reaches its maximum (72\,{\textmu}s$^{-1}$) at $T \sim$ 10\,K, i.e., well 
above $T_\mathrm{N}$, this implies significant spin fluctuations in the PM  state of SmAlGe. A similar behavior has been previously
observed in CeAlGe, where spin fluctuations in the PM state
are stabilized by the band topology~\cite{Drucker2023}. Upon further
lowering the temperature, both $\lambda^\mathrm{L}_1$
and $\lambda^\mathrm{L}_2$ decrease, reaching 0.25 and 1.7\,{\textmu}s$^{-1}$ at 1.5\,K, respectively.
To conclude, spin fluctuations in the PM state are more
significant in SmAlGe than in NdAlGe. Conversely, in the AFM/FIM state,
NdAlGe shows more significant spin fluctuations than SmAlGe. Future calculations of
the muon-stopping sites might be helpful to better appreciate the
differences between NdAlGe and SmAlGe.

%%%%%%%%%%%%%%%%%%%%%%%%%%%%%%%%%%%%%
\begin{table}[!bht]
	\centering
	\caption{Internal-field parameters of NdAlGe and SmAlGe single crystals, 
		as obtained from ZF-{\textmu}SR measurements. Here, $\Delta_i$ (= $\lambda^\mathrm{T}_i/\gamma_\mu$) represents the HWHM of the field distribution at 1.5\,K.}\label{tab:1}
	\begin{ruledtabular}
		\begin{tabular}{lcl}
			Parameter               & NdAlGe          &  SmAlGe \\
			$B^\mathrm{int}_{1}$(0) & 314(24) mT      &  97(1) mT \\
			$\gamma_1$              & 1.7(9)          &  1.56(7) \\
			$\delta_1$              & 0.41(9)         &  0.57(1) \\
			$\Delta_1$              & 135(14) mT      &  36(2) mT  \\
			\midrule
			$B^\mathrm{int}_{2}$(0) & 474(9) mT       &  191(12) mT \\
			$\gamma_2$              & 2.2(3)          &  1.8(6) \\
			$\delta_2$              & 0.62(4)         &  0.8(2) \\
			$\Delta_2$              & 62(6) mT        &  69(13) mT  \\
		\end{tabular}
	\end{ruledtabular}
\end{table}%%%%%%%%%%%%%%%%%

%%%%%%%%%%%%%%%%%
\begin{figure}[!htp]
	\centering
	\includegraphics[width=0.45\textwidth]{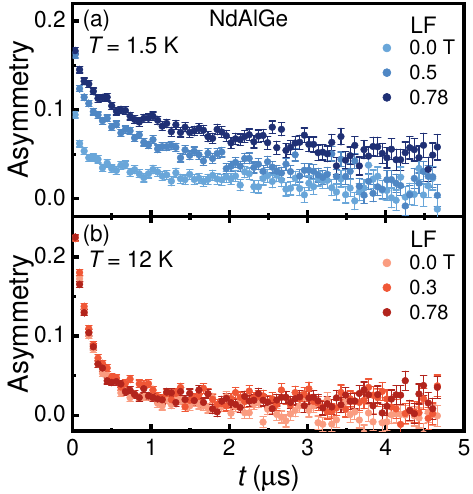}
	\caption{(Color online) LF-{\textmu}SR spectra collected at 1.5\,K (a) (FIM state) and 12\,K (b) (PM state) in applied magnetic fields up to 0.78\,T in NdAlGe single crystals. The spectra were collected in a longitudinal muon-spin configuration ($\boldsymbol{P}_\mu$ $\parallel$ $\boldsymbol{S}_\mu$). The applied magnetic field is parallel to the muon-spin direction. In either case, no appreciable decoupling of muon spins with field can be identified.}
	\label{fig:15}
\end{figure}
%%%%%%%%%%%%%%%%%

To further investigate the spin fluctuations, we performed LF-{\textmu}SR measurements in NdAlGe, while we expected similar features in SmAlGe. As shown in Fig.~\ref{fig:15}, 
the LF-{\textmu}SR spectra were collected under various longitudinal fields up to 0.78\,T at temperatures well inside the FIM state (1.5\,K) and in the PM state (12\,K) of NdAlGe.  
In the magnetically ordered state, the fast drop of the zero-field {\textmu}SR asymmetry reflects a very fast muon-spin depolarization in the first tenths
of {\textmu}s [see also ZF-{\textmu}SR in Fig.~\ref{fig:12}(a)]. 
As the applied longitudinal field overcomes the internal field ($\sim$0.48\,T), the static magnetism (transverse fraction of the {\textmu}SR signal) starts to decouple from the muon spin [see Eq.~\eqref{eq:3}], here reflected in an up-shift of 
the long-time {\textmu}SR spectra. However, a 0.78\,T longitudinal field has negligible effects on the dynamic part (i.e., the shape of relaxation curve), 
implying that spin fluctuations persist deep inside the FIM state of NdAlGe, consistent with the $\lambda^\mathrm{L}_i(T)$ in Fig.~\ref{fig:13}(b)-(c).
In the PM state, as shown in Fig.~\ref{fig:15}(b), the {\textmu}SR spectrum in a 0.78\,T longitudinal field is almost identical
to that collected in a zero-field condition, suggesting that muon spins cannot be decoupled either. Therefore, spin fluctuations are robust against the external field in both the FIM and PM states of NdAlGe. Similar {\textmu}SR results have been reported in other rare-earth-based materials, e.g., Eu(Al,Ga)$_4$ and EuCd$_2$As$_2$, where the strong spin
fluctuations play a significant role in their topological properties in both the real and the momentum spaces~\cite{Zhu2022,Ma2019,xu2021}. It is worth mentioning that
topological Hall resistivity is observed in the $\sim 0.2$--2\,T  field range in NdAlGe at $T < T_\mathrm{N}$, while it is absent in SmAlGe.
A possible reason might be the much weaker spin fluctuations deep inside
the magnetically ordered state of SmAlGe, while 
they remain significantly
stronger in NdAlGe. For example, at 1.5\,K,
$\lambda^\mathrm{L}$ $\sim 36$\,{\textmu}s$^{-1}$ in NdAlGe vs
$\sim 1.7$\,{\textmu}s$^{-1}$ in SmAlGe. 
This comparison indicates clearly that spin fluctuations
may be crucial to the onset of topological transport
properties in the magnetic $RE$Al(Si,Ge) Weyl semimetals.

\section{Discussion}\label{sec:dis}

First, we discuss the magnetic phase diagram of NdAlGe. According to 
neutron diffraction studies, below $T_\mathrm{N}$, NdAlGe
hosts a multi-$\boldsymbol{k}$ magnetic structure~\cite{Yang2023,Dhital2023},
with propagation vectors
$\boldsymbol{k}_\mathrm{AFM1} = (2/3 + \delta, 2/3 + \delta, 0)$,
$\boldsymbol{k}_\mathrm{AFM2} = (1/3 - \delta, 1/3 - \delta, 0)$, and 
$\boldsymbol{k}_\mathrm{FM}   = (3\delta, 3\delta, 0)$, where $\delta$
is a measure of incommensurability (e.g., $\delta$ $\sim$0.01 at 6\,K),
which becomes zero in the commensurate FIM state. 
There are two Nd$^{3+}$ ions
in the magnetic unit cell, whose moments are aligned antiparallel
along the $c$ axis with a small helical canting within the $ab$ plane,
creating a down-up-up ferrimagnetic
spin structure along the $\boldsymbol{k}$-vector direction~\cite{Yang2023,Dhital2023}. 
As shown in Fig.~\ref{fig:6}, for $H \parallel c$, four different
magnetic phases can be identified below $T_\mathrm{N}$, including:
i) FIM1 at $H < H_{c1}$, down-up-up configuration with up-down-down stripe domains; ii) FIM2 at $H_{c1} < H < H_{c2}$, down-up-up ordering state; iii)  FIM3 at $H_{c2} < H < H_{c3}$, canted down-up-up state, where the canting angle is largely increased; and finally iv) PFM at $H > H_{c3}$, polarized up-up-up state~\cite{Yang2023,Dhital2023,Zhao2022}. Interestingly, the topological Hall resistivity appears only in the
$H \parallel c$ FIM2 phase for $H_{c1} < H < H_{c2}$ [see details in Fig.~\ref{fig:7}(b)].
A similar phase diagram has been identified also for $H \parallel ab$.
Here, considering that neither $M(H)$ nor $\rho_{xy}(H)$ show any anomalies, 
the $H_{c1}$ and $H_{c2}$ might correspond to the critical fields
resulting from a small canting of Nd$^{3+}$ moments.

Second, we discuss the possible origin of the topological Hall effect in NdAlGe single crystals.
In most topological magnetic materials, the THE is attributed to the Berry curvatures created by the topological spin textures, such as chiral domain walls or skyrmions. 
NdAlGe exhibits a similar magnetic structure to CeAlGe and NdAlSi, both described by a
mul\-ti\--$\boldsymbol{k}$ structure~\cite{puphal2020,Gaudet2021}.
The THE can be generated by the chiral domain walls or by the
scattering of Weyl fermions through the domain walls. Such scenarios
have been proposed to explain the topological transport properties
of CeAlSi~\cite{piva2023}, where the evolution of domains
and domain walls with temperature has been experimentally
verified via the magneto-optical Kerr effect~\cite{sun2021}.
Since, in NdAlGe, neutron scattering has confirmed the presence
of stripe domains at low fields and a dozen of Weyl nodes
near the Fermi level have been theoretical identified~\cite{Yang2023}, it is
reasonable to speculate that domain walls play an important role
in the topological transport properties of NdAlGe as well. 

%It is
%worth mentioning that the large anomalous Hall effect in NdAlGe
%is due to the intrinsic Berry curvature generated also \tcr{by the Weyl
	%nodes near the Fermi level!}.
% Please note that, in the original paper Yang et al, suggest that
% Weyl-mediated magnetism prevails in the NdAlX Weyl semimetals. Yet,
% transport properties - including AHE - are affected by the material-specific
% extrinsic effects such as disorder, despite the presence of a prominent Berry curvature.

The breaking of inversion symmetry in the $RE$Al(Si,Ge) family
can give rise to Dzya\-lo\-shin\-skii\--Moriya interactions,
which compete with RKKY interactions and represent one of the
key ingredients of topological spin textures. 
%represent one of the key ingredients of topological spin textures. 
In CeAlGe, the incommensurate multi-$\boldsymbol{k}$ magnetic structure can host skyrmion pairs, which account for the  presence of topological transport properties~\cite{puphal2020,piva2023-1}. 
Since NdAlGe shows a similar magnetic phase diagram to CeAlGe, we expect
that also in our case THE might be attributed to the presence of magnetic
skyrmions. In general, the manifestation of THE requires a noncollinear or a noncoplanar magnetic structure, which leads to a finite scalar spin chirality $\chi_{ijk}$ = $\boldsymbol{S}$$_i$ $\cdot$ ($\boldsymbol{S}$$_j$ $\times$ $\boldsymbol{S}$$_k$) in real space.
Although neutron diffraction suggests an almost collinear multi-$\boldsymbol{k}$ magnetic structure
along the $c$ axis (with only a small $ab$-plane component in the
low-field region of NdAlGe~\cite{Yang2023,Dhital2023}), whether
a noncollinear or noncoplanar structure can be induced by
an externally applied magnetic field remains an open question.
Moreover, recently the Co$_7$Zn$_7$Mn$_6$ chiral magnet was found
to host a skyrmion phase far below the magnetic ordering temperature.
Here, spin fluctuations are believed to be the key for stabilizing
the magnetic skyrmions~\cite{ukleev2021}.
Further, in Ir/Fe/Co/Pt heterostructures, there is evidence that
chiral spin fluctuations can enhance the THE near the transition
regime between isolated skyrmions and skyrmion lattices~\cite{raju2021}.
Similar to the above cases, our {\textmu}SR results reveal
significant spin fluctuations in the magnetically ordered state
of NdAlGe, which are robust against the external magnetic field.
This fact suggests that spin fluctuations are a crucial factor
for understanding the origin of THE and of possible skyrmion phases in
NdAlGe. Deciding which scenario, chiral domain walls or skyrmions,
may account for the observed THE in NdAlGe, requires future field-dependent
neutron diffraction measurements.

Finally, we discuss the possible reasons for the absence of THE (or AHE)
in SmAlGe. Such an absence can be either extrinsic or intrinsic. The $RE$Al(Si,Ge)
single crystals grown by the Al-flux method (see details in section~\ref{ssec:structure}), tend to be
rich in Al %more vacancies on the Ge or Si sites 
than crystals grown using the floating-zone method, where the Ge or Si atoms in the former cases are partially substituted by Al atoms~\cite{Yang2023,Puphal2019,kikugawa2024}.
The electronic properties of $RE$Al(Si,Ge) are extremely sensitive to small stoichiometric variations, as already shown in the CeAlGe case~\cite{hodovanets2018,hodovanets2022,piva2023-1}.
Even small changes in chemical composition may shift the Fermi level and, thus, result in a significantly different Fermi surface and Berry curvatures. As a consequence, the THE present in CeAlGe and NdAlGe is absent in other crystals with almost identical chemical composition~\cite{Kikugawa2023,hodovanets2018}, which might also be the case for SmAlGe.
Moreover, since in $RE$Al(Si,Ge) the magnetic order is strongly
coupled to the Fermi pockets and Weyl nodes~\cite{Yang2023,Dhital2023,Gaudet2021}, different magnetic orders, including FM or AFM, or both can be present in the same material, depending on 
stoichiometric variations~\cite{hodovanets2018,hodovanets2022}. 
It seems that the topological transport properties of $RE$Al(Si,Ge)
are also sensitive to the sample quality and/or disorder~\cite{Yang2023}. 
Indeed, the higher the crystal quality [reflected in a large
residual resistivity ratio (RRR)], the larger the THE effect observed
in a single crystal of CeAlGe~\cite{piva2023-1}. Since our SmAlGe
has an estimated RRR of 1.2, this implies sizable disorder effects,
which might mask or even suppress possible topological features.

As for the intrinsic case, the magnetic order in SmAlGe is quite
robust against externally applied fields (see Fig.~\ref{fig:8}).
This reflects the stronger Sm magnetic interactions compared to the Ce, Pr, or Nd counterparts. In SmAlGe, the field-dependent magnetization $M(H)$ is linear in field up to 7\,T, indicating the absence of metamagnetic transitions and, thus, the stability of its AFM magnetic structure. At 7\,T, the SmAlGe magnetization is about 0.045~{\textmu}$_\mathrm{B}$,
i.e., only $\sim 6.3$\% of the saturation magnetization of free Sm$^{3+}$
ions (0.71{\textmu}$_\mathrm{B}$). Thus, if any metamagnetic transitions
exist in SmAlGe, they can occur only at higher magnetic fields. Second, if spin fluctuations are indeed crucial for the formation of topological spin textures in NdAlGe, then the absence of THE (or AHE) in SmAlGe is expected. According to our ZF-{\textmu}SR measurements (see Fig.~\ref{fig:13}), spin fluctuations in the magnetically ordered state of SmAlGe are significantly
weaker than in the NdAlGe case. Moreover, the competition between four-spin interactions and magnetic anisotropy has been proposed to explain the formation of skyrmions in the centrosymmetric tetragonal material GdRu$_2$Si$_2$~\cite{khanh2020}. Here, we find that,
while NdAlGe exhibits a strong magnetic anisotropy (see Fig.~\ref{fig:4}),
SmAlGe is almost isotropic, reflected in the identical field-dependent
magnetization for $H \parallel c$ and $H \parallel ab$ (see Fig.~\ref{fig:8}). The lack of magnetic anisotropy might also account for the absence of THE in SmAlGe. Despite
the possibility of SmAlGe to show THE or AHE,
we presume that most likely it fails to fulfill the right conditions.
Overall, the $RE$Al(Si,Ge) family represents an excellent platform for studying the interplay between band topology and magnetism, as well as the emergent topological transport properties.

%\vspace{7pt}
\section{\label{ssec:Sum}Conclusion}

In summary, by systematic electrical resistivity and magnetization measurements, we could establish the magnetic phase diagram of NdAlGe and SmAlGe single crystals for both $H \parallel c$ and $H \parallel ab$. 
The wTF-{\textmu}SR spectra reveal a magnetic volume fraction of about 90\% in both NdAlGe and SmAlGe, implying a good sample quality. While NdAlGe exhibits a
rich magnetic phase diagram, including multiple metamagnetic
transitions under applied field, the SmAlGe magnetic order
is very robust against externally applied field.
Interestingly, in NdAlGe, a topological Hall effect
appears below $T_\mathrm{N}$ between the first- and the second
metamagnetic transitions (i.e., $H_{c1} < H < H_{c2}$) for
$H \parallel c$, while it is absent for $H \parallel ab$, as well as in SmAlGe.  
The presence of THE in NdAlGe is most likely attributed to the topological
spin textures. By contrast, the absence of topological transport
properties in SmAlGe can be either extrinsic or intrinsic. 
According to our ZF-{\textmu}SR measurements, NdAlGe exhibits a more disordered internal field distribution than SmAlGe, reflected in a larger transverse muon-spin relaxation rate $\lambda^\mathrm{T}$ deep inside the magnetically ordered state, most likely related to its
multi-$\boldsymbol{k}$ magnetic structure. The vigorous spin fluctuations revealed by both ZF-{\textmu}SR
and LF-{\textmu}SR might be crucial for understanding the origin of THE and of possible topological spin textures in $RE$Al(Si,Ge) Weyl semimetals. Which scenario, chiral
domain walls or skyrmions, may account for the THE in NdAlGe,
requires future
field-dependent neutron diffraction measurements. It might be interesting
to study the magnetic properties of NdAlGe also via the {\textmu}SR technique
under higher magnetic fields, where the THE appears.\\
\begin{acknowledgments} 
This work was supported by the Natural Science Foundation
of Shanghai (Grant Nos.\ 21ZR1420500 and 21JC\-140\-2300), Natural Science
Foundation of Chongqing (Grant No.\ CSTB-2022NSCQ-MSX1678), National
Natural Science Foundation of China (Grant No. 12374105), Fundamental Research Funds for the Central Universities, and the
Schweizerische Nationalfonds zur F\"{o}r\-der\-ung der
Wis\-sen\-schaft\-lichen For\-schung (SNF) (Grant Nos.\ 200021\_169455 and 200021\_188706). 
Igor Plokhikh acknowledges support from Paul Scherrer Institute research grant (Grant No.\ 202101346).
We acknowledge the allocation of beam time at the Swiss muon source (GPS {\textmu}SR spectrometer).
\end{acknowledgments}

\appendix*
\section{}

The supporting information is available online at http://phys.scichina.com
and https://link.springer.com. The supplementary crystallographic data for
the materials mentioned in this paper, NdAlGe and SmAlGe, are also deposited
in the CCDC database with access numbers 2353094 and 2353095, respectively.
The supporting materials are published as submitted, without typesetting
or editing. The responsibility for scientific accuracy and content remains
entirely with the authors.

%\begin{footnotesize}
\bibliography{Mo3P_prb_R5.bib}
%\end{footnotesize}

\end{document}